\documentclass[aps,prl,twocolumn,superscriptaddress,showpacs]{revtex4}
\usepackage{graphicx,graphics,amssymb,amsmath,hyperref}

\begin{document}

\def\etal#1{, #1}
\def\tit#1{}
\def\jour#1{#1}

\title{Spin-chain system as a tunable simulator of frustrated planar magnetism}

\author{M. Klanj\v{s}ek}
\email{martin.klanjsek@ijs.si}
\affiliation{Jo\v{z}ef Stefan Institute, Jamova 39, 1000 Ljubljana, Slovenia}
\affiliation{EN-FIST Centre of Excellence, Dunajska 156, 1000 Ljubljana, Slovenia}
\affiliation{Laboratoire National des Champs Magn\'etiques Intenses, LNCMI - CNRS (UPR3228), UJF, UPS and INSA, BP 166, 38042 Grenoble Cedex 9, France}

\author{M. Horvati\'c}
\author{C. Berthier}
\author{H. Mayaffre}
\affiliation{Laboratoire National des Champs Magn\'etiques Intenses, LNCMI - CNRS (UPR3228), UJF, UPS and INSA, BP 166, 38042 Grenoble Cedex 9, France}

\author{E. Can\'evet}
\affiliation{Institut Laue-Langevin, BP 156, 38042 Grenoble Cedex 9, France}
\affiliation{SPSMS, UMR-E 9001, CEA-INAC / UJF, Laboratoire Magn\'etisme et Diffraction Neutronique, 38054 Grenoble Cedex 9, France}

\author{B. Grenier}
\affiliation{SPSMS, UMR-E 9001, CEA-INAC / UJF, Laboratoire Magn\'etisme et Diffraction Neutronique, 38054 Grenoble Cedex 9, France}

\author{P. Lejay}
\affiliation{Institut N\'eel - CNRS, UJF, 38042 Grenoble Cedex 9, France}

\author{E. Orignac}
\affiliation{LPENSL CNRS UMR 5672, 69364 Lyon Cedex 7, France}

\date{\today}

% 75.10.Hk Classical spin models
% 75.10.Jm Quantized spin models, including quantum spin frustration
% 75.10.Pq Spin chain models
% 75.25.-j Spin arrangements in magnetically ordered materials (including neutron and spin-polarized electron studies, synchrotron-source x-ray scattering, etc.)
% 75.30.Kz Magnetic phase boundaries (including classical and quantum magnetic transitions, metamagnetism, etc.)
% 75.40.Cx Static properties (order parameter, static susceptibility, heat capacities, critical exponents, etc.)
% 76.60.-k Nuclear magnetic resonance and relaxation

\pacs{75.10.Jm, 75.10.Pq, 75.25.-j, 76.60.-k}

\begin{abstract}
At low temperatures, weakly coupled spin chains develop a magnetic order that reflects the character of gapless spin fluctuations along the chains. Using nuclear magnetic resonance, we identify and characterize two ordered states in the gapless region of the antiferromagnetic, Ising-like spin-chain system BaCo$_2$V$_2$O$_8$, both arising from the incommensurate fluctuations along the chains. They correspond to the columnar and ferromagnetic ordered states of the frustrated $J_1$-$J_2$ spin model on a square lattice, where the spins are encoded in original spin chains. As a result of field-dependent incommensurate fluctuations and frustrated interchain interaction, $J_1$ can be tuned continuously with the magnetic field, and its value with respect to a fixed $J_2$ selects the ordered state. Spin-chain systems can thus be used as tunable simulators of frustrated planar magnetism.
\end{abstract}

\maketitle

The study of important models in condensed-matter physics increasingly relies on the concept of quantum simulation. This concept is largely associated with the gases of cold atoms, whose well-defined and highly tunable hamiltonians can be mapped onto many model hamiltonians of interest~\cite{Bloch_2008}. Cold atoms in optical lattices allowed the first simulation of the quantum phase transition from a superfluid to a Mott insulator~\cite{Greiner_2002}, and, recently, also the simulation of ordered magnetic states on a lattice~\cite{Struck_2011,Simon_2011}. Scarce difficulties with cold atoms, related to the fact that they are gases, are absent in quantum antiferromagnets, which also feature well-defined hamiltonians. Among them, those that contain dimers of quantum spins can host bosonic excitations. As the density of bosons can simply be driven by the applied magnetic field, these systems have proven as convenient simulators of Bose-Einstein condensates~\cite{Nikuni_2000,Ruegg_2003,Giamarchi_2008} and Tomonaga-Luttinger liquids (TLLs)~\cite{Klanjsek_2008}.

In exploring the field-temperature phase diagram of the antiferromagnetic, Ising-like spin-chain system BaCo$_2$V$_2$O$_8$~\cite{Wichmann_1986,He_2005_1,He_2005_2,Kimura_2006,Kimura_2007,Kimura_2008_1,Kimura_2008_2,Lejay_2011}, we discovered a new low-temperature ordered state above $B_p=8.6$~T. The state borders on the recently identified, novel incommensurate (IC) ordered state between $B_c=3.9$~T~\cite{Kimura_2008_1,Kimura_2008_2} and $B_p$, shares its quantum origin, but exhibits a different order. We show that if we map the configuration of spins in each chain onto a classical spin, the two ordered states correspond to the columnar and ferromagnetic ordered states of the frustrated $J_1$-$J_2$ spin model on a square lattice~\cite{Chandra_1988,Schulz_1992,Melzi_2000,Shannon_2004}. Remarkably, the effective exchange coupling $J_1$ between the nearest-neighboring (NN) chains (i.e., classical spins) can be continuously tuned by the applied magnetic field from {\em ferromagnetic} to {\em antiferromagnetic}, and we extract this field dependence. The value of $J_1$ with respect to a fixed exchange coupling $J_2$ between the next-nearest-neighboring (NNN) chains (i.e., classical spins) then selects the most favorable ordered state~\cite{Shannon_2004}. This is a mechanism for the transition between the two ordered states at $B_p$. From the practical point of view, the exchange coupling scheme encountered in BaCo$_2$V$_2$O$_8$ allows a simple {\em quantitative} control of the exchange interaction between the effective classical spins, extending the simulation possibilities of quantum antiferromagnets to the domain of planar magnetism. This is all the more valuable as a direct control of exchange interactions in known materials, by changing the applied pressure~\cite{Ruegg_2008} or the chemical composition~\cite{Jia_2011}, is only {\em qualitative}.

We start by explaining the principle of the tuning mechanism. It is based on two ingredients: (i) chains hosting a type of spin fluctuations that can be manipulated by the magnetic field, and (ii) a frustrated interaction between pairs of chains. The tuning mechanism applies to chains formed by $S=1/2$ spins that are coupled by an anisotropic, Ising-like antiferromagnetic exchange interaction~\cite{Yang_1966,Haldane_1980,Hikihara_2004,Suga_2008}. When the magnetic field points along the exchange-anisotropy axis $z$, a single chain exhibits gapless spin fluctuations in a certain field range, thus realizing a TLL~\cite{Giamarchi_2004}. In this range, i.e., between the critical fields $B_c$ and $B_s$, a uniform magnetization $m_z$ per spin induced by the field $B$ grows monotonously with $B$ from $0$ to the saturated value of $1/2$. In the lower part of this range, gapless fluctuations are longitudinal (i.e., they involve spin components along the field) and appear at the field-dependent IC wavevector $2k_F(B)=\pi [1-2m_z(B)]$ along the chain~\cite{Okunishi_2007}. Mapping the spin chain onto a one-dimensional (1D) system of spinless fermions using the Jordan-Wigner transformation~\cite{Giamarchi_2004}, this equation simply relates the filling of the fermion band $1/2-m_z$ with the wavevector $2k_F$ connecting both points of the Fermi surface. Gapless fluctuations prevent an isolated chain from ordering even at zero temperature. However, in weakly coupled chains, a 3D magnetic order develops at low enough temperatures, whose type reflects the character of fluctuations: spins develop an ordered component along the field (on top of the uniform $m_z$) in the form of spin-density waves (SDWs) with an IC wavevector $2k_F(B)$ along the chains~\cite{Okunishi_2007}. The main point here is that the wavevector of SDWs is controlled by the field, varying from $\pi$ to $0$ between the critical fields. This is our first ingredient.

\begin{figure}
\includegraphics[width=1\linewidth]{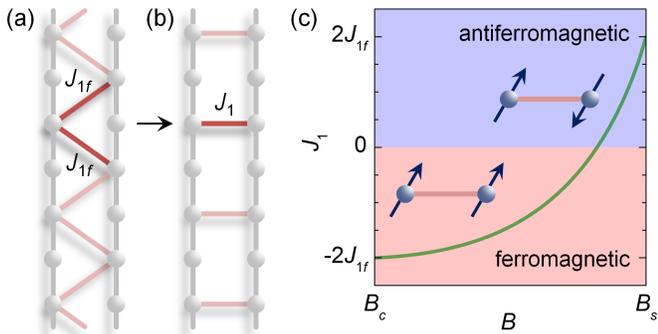}
\caption{
(color online) (a),(b) A frustrated pair of antiferromagnetic $J_{1f}$ couplings in the zigzag scheme sums up to the effective coupling $J_1(B)$ given by Eq.~(\ref{eq_J1s}), if the chains are hosting ordered SDWs with the wavevector $2k_F(B)$. (c) $J_1(B)$ based on the numerically calculated $m_z(B)$ for exchange anisotropy $\Delta=2.17$ in the range between the critical fields $B_c$ and $B_s$. SDWs are mapped onto the classical spins (arrows) that assume a ferromagnetic (antiferromagnetic) arrangement for a negative (positive) $J_1(B)$.
}
\label{fig1}
\end{figure}

To make the energy of the exchange-coupled SDWs depend on their wavevector, and hence on the magnetic field, pairs of SDWs should interact in a frustrated manner. This is our second ingredient. Namely, in the case of a non-frustrated interaction based on the couplings $J_1$ [Fig.~\ref{fig1}(b)], the energy of the coupled pair of SDWs is proportional to $\frac{1}{2}J_1\cos(\varphi_1-\varphi_2)$, where $\varphi_1$ and $\varphi_2$ are their phases. There is no wavevector dependence. In contrast, for the simplest case of a frustrated interaction based on a zigzag scheme with the couplings $J_{1f}$ [Fig.~\ref{fig1}(a)], the coupling energy is proportional to $J_{1f}\cos(2k_F)\cos(\varphi_1-\varphi_2)$. The wavevector-dependent $\cos(2k_F)$ factor simply reflects a one-site offset between the coupled sites in the respective chains. A comparison of the cases displayed in Figs.~\ref{fig1}(a) and (b) shows that each frustrated pair of $J_{1f}$ couplings sums up to the field-dependent effective coupling
\begin{equation}\label{eq_J1s}
	J_1(B)=2J_{1f}\cos[2k_F(B)].
\end{equation}
Furthermore, due to the $\cos(\varphi_1-\varphi_2)$ factor the energy of the coupled SDWs adopts the standard form of the coupling energy between the classical planar spins with orientations $\varphi_1$ and $\varphi_2$. The phases of SDWs thus encode the orientations of the effective classical spins interacting with the coupling $J_1(B)$ given by Eq.~(\ref{eq_J1s}). This coupling can be tuned continuously by the field $B$ in the range between the critical fields, where it even changes sign [Fig.~\ref{fig1}(c)], which is an unexpected feature. Although the individual interchain couplings $J_{1f}$ are antiferromagnetic, $J_1(B)$ can be tuned from ferromagnetic to antiferromagnetic. Accordingly, the effective classical spins flip from a parallel to an antiparallel arrangement [Fig.~\ref{fig1}(c)], implying that the original SDWs switch from being in phase to being in opposite phase.

\begin{figure*}
\includegraphics[width=1.0\linewidth]{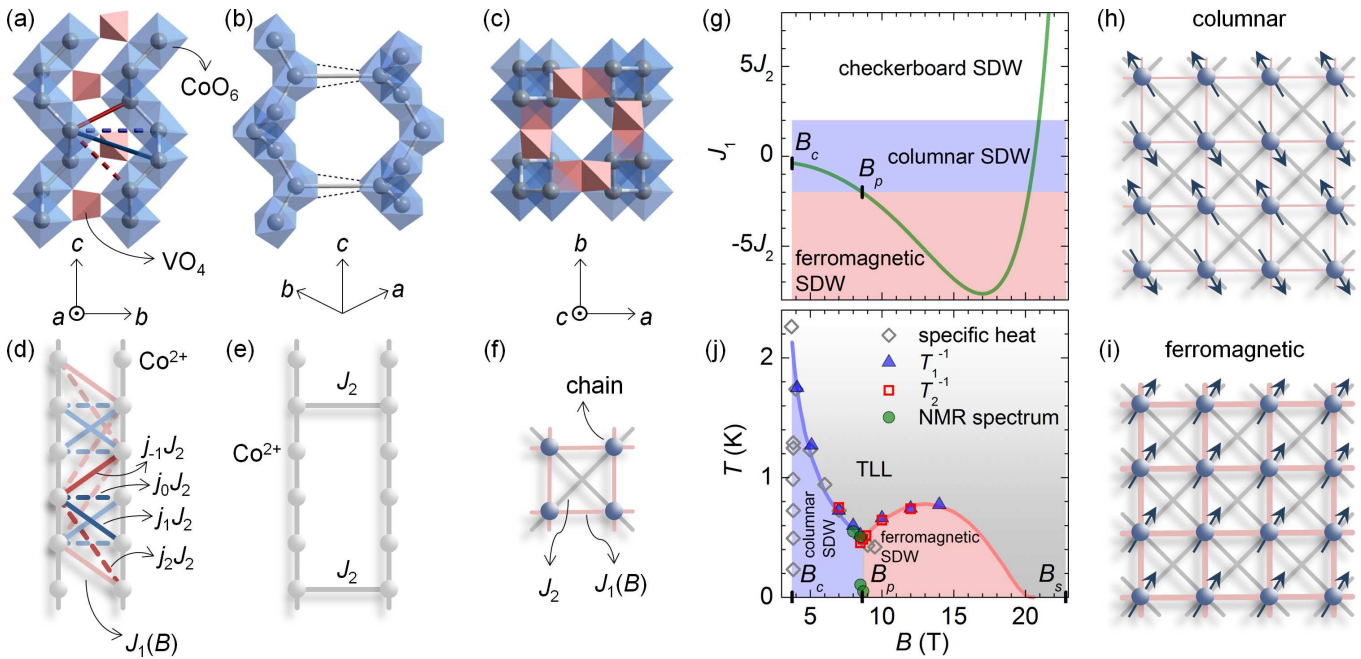}
\caption{
(color online) (a)-(c) Crystal structure of BaCo$_2$V$_2$O$_8$ contains edge-sharing CoO$_6$ octahedra forming the chains along $c$. The dominant exchange interaction between Co$^{2+}$ spins runs along the chains. Much smaller exchange interaction between Co$^{2+}$ spins in neighboring chains is mediated by the O-O bridges. Corresponding schemes of interchain exchange couplings are shown in (d)-(f). (d) Couplings of four different sizes [highlighted, also shown in (a)] between the NN chains combine into eight frustrated pairs, which sum up to the tunable effective coupling $J_1(B)$. (e) Non-frustrated interaction between the NNN chains is based on couplings $J_2$. (f) Chains form a frustrated square lattice in the $a$-$b$ plane, where $J_1(B)$ competes with $J_2$. (g) $J_1(B)$ given by Eq.~(\ref{eq_J1c}) calculated for $j_0=1.0$, $j_{-1}+j_1=3.4$, $j_2=2.3$ [all defined in (d)] and exchange anisotropy $\Delta=2.17$ between the critical fields $B_c=3.9$~T and $B_s=22.7$~T. The value of $J_1(B)$ with respect to $J_2$ selects the ordered state. (h),(i) Sketches of the columnar and ferromagnetic ordered states with SDWs mapped onto the classical spins (arrows). (j) Field-temperature phase diagram obtained from the NMR data shown in Figs.~\ref{fig4}(a) and (b) for a field parallel to $c$. It agrees with the specific heat data available up to $\sim\!\!B_p=8.6$~T~\cite{Kimura_2008_1}. Lines are calculated phase boundaries based on $J_1(B)$ [plotted in (g)] between the TLL state and the ordered states.
}
\label{fig2}
\end{figure*}

In the following we demonstrate that such a tuning mechanism is realized in BaCo$_2$V$_2$O$_8$, which contains parallel chains of magnetic Co$^{2+}$ ions, each carrying an effective anisotropic $S=1/2$ spin~\cite{Kimura_2006,Kimura_2007}. The chains run along the $c$~axis and form a square lattice in the tetragonal $a$-$b$ plane [Figs.~\ref{fig2}(a)-(f)]. The exchange-anisotropy axis coincides with the chain direction, which is also chosen as the field direction. Interaction between the NNN chains (neighbors along the bisectors of $a$ and $b$) is not frustrated and is given by the field-independent coupling $J_2$ [Figs.~\ref{fig2}(b) and (e)]. A coupling scheme between the NN chains (neighbors along $a$ or $b$) is frustrated, but more complicated than the simple zigzag scheme. It contains the couplings of four different sizes [expressed by the weights $j_{-1}$, $j_0$, $j_1$ and $j_2$ relative to $J_2$ in Fig.~\ref{fig2}(d)], which combine into eight frustrated pairs within four chain units [Fig.~\ref{fig2}(d)]. As each pair contributes a term similar to Eq.~(\ref{eq_J1s}), together they amount to the effective coupling
\begin{equation}\label{eq_J1c}
	J_1(B)=4J_2\Bigl[j_0+(j_{-1}+j_1)\cos[2k_F(B)]+j_2\cos[4k_F(B)]\Bigr].
\end{equation}
The non-monotonous shape of $J_1(B)$ displayed in Fig.~\ref{fig2}(g) is generic, while the precise values depend on $j_0$, $j_{-1}+j_1$ and $j_2$.

According to Fig.~\ref{fig2}(f), in the picture with SDWs mapped onto the classical spins, a coupled chain system in BaCo$_2$V$_2$O$_8$ maps onto the frustrated $J_1$-$J_2$ spin model on a square lattice. NN spins in this model interact with the tunable coupling $J_1(B)$ and NNN spins with the field-independent coupling $J_2$. These two couplings compete to determine the most favorable spin arrangement. The effective classical spins are free to rotate in a plane and, for the arrangement described by the wavevector ${\bf k}=(k_x,k_y)$, the energy of the model reads~\cite{Shannon_2004}
\begin{equation}\label{eq_enJ1J2}
	\varepsilon({\bf k})\propto \frac{J_1(B)}{2}(\cos{k_x}+\cos{k_y}) + J_2\cos{k_x}\cos{k_y}.
\end{equation}
Depending on the relative values of $J_1(B)$ and $J_2$, this energy is minimized in one of the three ground states: ferromagnetic with ${\bf k}=0$ for $J_1(B)<-2J_2$ [Fig.~\ref{fig2}(i)], columnar with ${\bf k}=(\pi,0)$ or $(0,\pi)$ for $-2J_2<J_1(B)<2J_2$ [Fig.~\ref{fig2}(h)], or checkerboard with ${\bf k}=(\pi,\pi)$ for $J_1(B)>2J_2$ (not shown)~\cite{Shannon_2004}. In the original picture, ${\bf k}$ describes the arrangement of the phase shifts of SDWs in the $a$-$b$ plane. Finally, as we are free to tune $J_1(B)$ in accordance with Fig.~\ref{fig2}(g), covering all three ground states, we are in possession of a simulator of frustrated magnetism on a square lattice.

\begin{figure}
\includegraphics[width=1\linewidth]{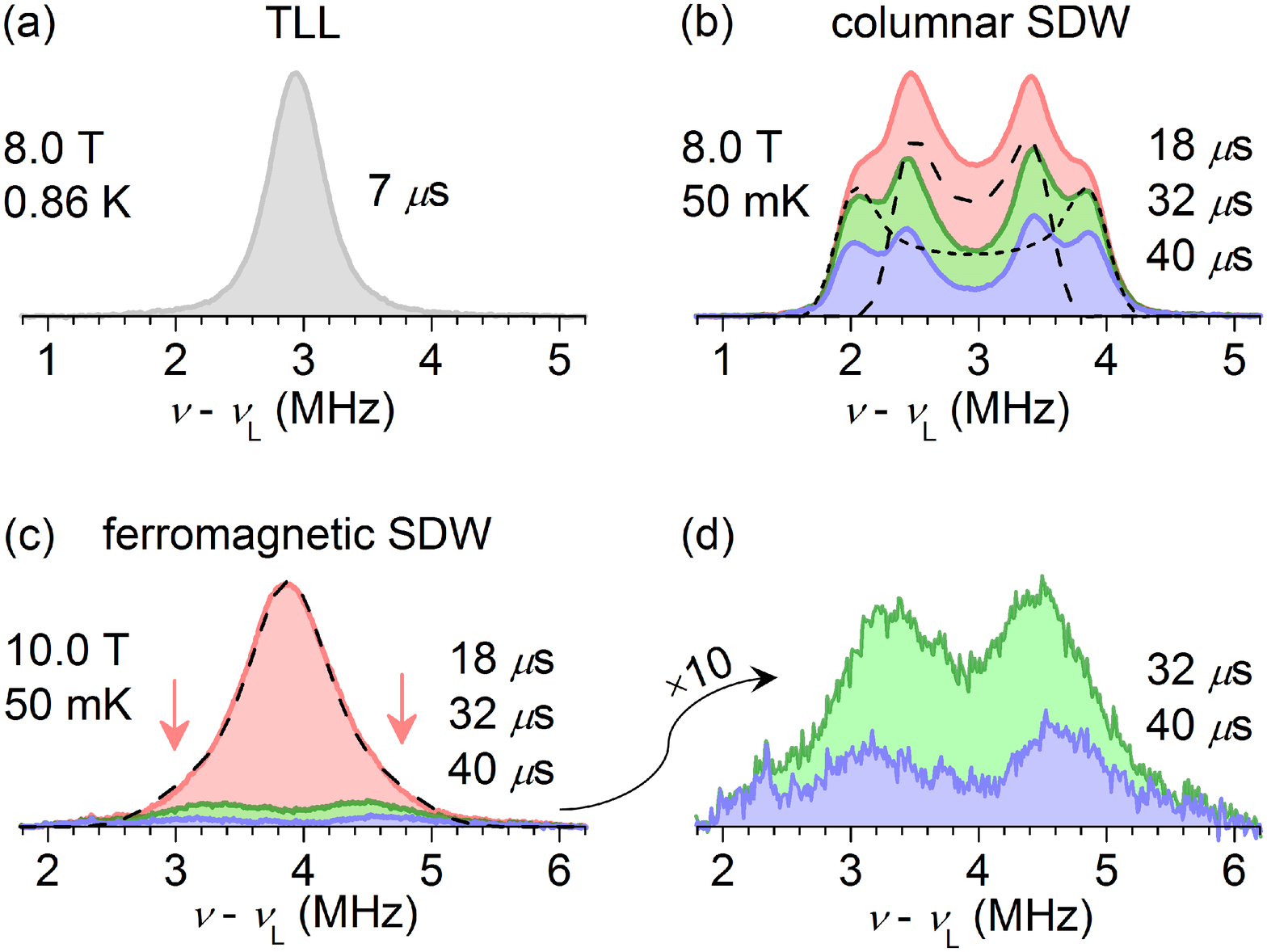}
\caption{
(color online) Representative $^{51}$V NMR spectra of various states (for selected field and temperature values) are recorded in a field parallel to $c$ after different durations of $T_2$ relaxation (given). (a) In the TLL state the spectrum exhibits a single peak. (b) In the columnar SDW state, the unrelaxed spectrum is reproduced by two characteristic U-shaped components of the same area. They are generated by two sets of V sites, located in between the chains hosting SDWs that are either in phase (dashed line) or in opposite phase (dotted line). The $T_2$ relaxation preserves the shape of the spectrum. (c) In the ferromagnetic SDW state, the unrelaxed spectrum is reproduced by a distribution of U-shaped components generated by a single set of V sites (dashed line). A distribution based on the 2D sinusoidal modulation of the SDW amplitude in the $a$-$b$ plane results in a spectrum with a large central weight and pronounced wings (indicated by arrows). A longer $T_2$ relaxation suppresses the narrow U-shaped components and reveals the broad ones. (d) Broad components from (c) scaled up $10$-times.
}
\label{fig3}
\end{figure}

We check this picture experimentally by extracting the spin arrangement in various states from the corresponding $^{51}$V nuclear magnetic resonance (NMR) spectra~\cite{appendix}. At temperatures above the ordered states, throughout the TLL state, the spectrum exhibits a single peak [Fig.~\ref{fig3}(a)] generated by a single crystallographic V site located in between the NN chains [Figs.~\ref{fig2}(a) and (c)]. This symmetry breaks at the transition to one of the ordered states, and the spectrum develops features. The appearance of the sinusoidal SDW breaks the translational symmetry along the chains and leads to the characteristic U-shaped spectrum~\cite{Blinc_1981}. Between $B_c=3.9$~T and $B_p=8.6$~T, the spectrum develops two U-shaped components [Fig.~\ref{fig3}(b)], a fingerprint of the columnar SDW state. The components are generated by two sets of V sites arising from the broken symmetry between the $a$ and $b$ directions. Above $B_p$ and up to at least $15$~T, the highest field in our experiment, the spectrum consists of a distribution of U-shaped components, which are revealed during the transverse (i.e., $T_2$) relaxation [Figs.~\ref{fig3}(c) and (d)]. The distribution is generated by a single set of V sites, implying that the symmetry between $a$ and $b$ directions is restored. This suggests that either the ferromagnetic or the checkerboard SDW state is realized above $B_p$. Small linewidth of the spectrum is compatible with the ferromagnetic SDW state~\cite{appendix}. The spectrum is perfectly reproduced by a distribution that corresponds to the 2D sinusoidal modulation of the SDW amplitude in the $a$-$b$ plane [Fig.~\ref{fig3}(c)].

\begin{figure*}
\includegraphics[width=1.0\linewidth]{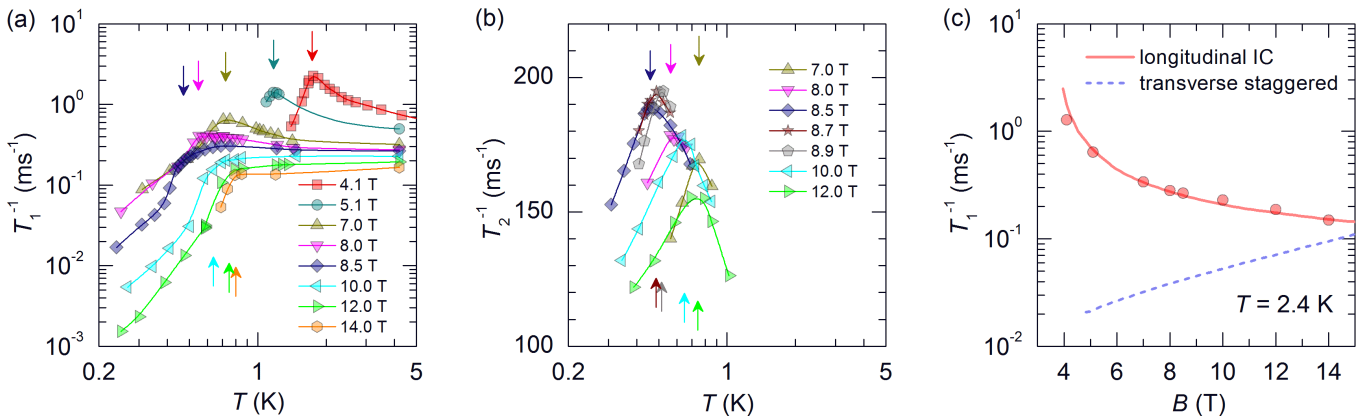}
\caption{
(color online) (a),(b) $^{51}$V longitudinal and transverse relaxation rates, $T_1^{-1}$ and $T_2^{-1}$, as a function of temperature $T$ in different magnetic fields parallel to $c$. Arrows mark the extracted temperatures of transition from the ordered SDW state to the TLL state. Solid lines are guides for the eye. (c) Interpolated $T_1^{-1}$ from (a) as a function of the field $B$ at $2.4$~K (in the TLL state) is compared to the TLL prediction for two types of spin fluctuations.
}
\label{fig4}
\end{figure*}

The field dependence of $J_1$ is deduced from the measured shape of the phase boundary between the ordered SDW states and the TLL state [Fig.~\ref{fig2}(j)]. Transition temperatures in different fields are obtained from the temperature dependence of the $^{51}$V longitudinal and transverse relaxation rates, $T_1^{-1}$ and $T_2^{-1}$~\cite{appendix}. In particular, the divergent behavior of $T_1^{-1}(T)$ and $T_2^{-1}(T)$ on increasing temperature $T$ is a signature of the second-order phase transition [Figs.~\ref{fig4}(a) and (b)]. The phase boundary changes the slope at exactly $B_p$, on the border between the two SDW states. In the columnar SDW state, Eq.~(\ref{eq_enJ1J2}) yields the coupling energy $\varepsilon\propto -J_2$, while in the ferromagnetic SDW state it is $\varepsilon\propto J_1(B)+J_2$. The average coupling of the chain to its neighbors in the two ordered SDW states is thus
\begin{equation}\label{eq_Jp}
	J'(B)=\left\{
		\begin{array}{ll}
			J_2 & {\rm if}\; B < B_p \\
			-J_1(B)-J_2 & {\rm if}\; B \geq B_p
		\end{array}
	\right.,
\end{equation}
which is continuous at $B_p$ because $J_1(B_p)=-2J_2$. For weakly interacting chains, the interchain coupling $J'$ can be treated in the mean-field approximation~\cite{Klanjsek_2008,Okunishi_2007,Schulz_1996}. This basically implies that the transition temperature scales as a field-dependent power of $J'$. Using $J'(B)$ given by Eq.~(\ref{eq_Jp}), the measured phase boundary is perfectly reproduced with the choice of $J_1(B)$ plotted in Fig.~\ref{fig2}(g) and the interchain coupling of $J_2/k_B=41$~mK ($k_B$ is the Boltzmann constant), which is indeed much smaller than the intrachain coupling of $65$~K~\cite{Kimura_2006,Kimura_2007}. Whereas the field-independent $J'(B)$ below $B_p$ implies a negative slope of the boundary (already known before our work~\cite{Okunishi_2007,Kimura_2008_1}), the slope inversion above $B_p$ is due to $J'(B)$ starting to grow steeply enough with field due to an increasing $-J_1(B)$ [Fig.~\ref{fig2}(g)]. As the mean-field expression for the transition temperature assumes a uniform SDW amplitude in the $a$-$b$ plane, a perfect agreement with the experimental points suggests that the observed 2D amplitude modulation in the ferromagnetic SDW state is nearly uniform, i.e., is long wavelength. We also note that the extracted $J_1(B)$ in Fig.~\ref{fig2}(g) predicts the existence of the checkerboard SDW state above $20$~T, which remains to be studied in high-field experiments.

Finally, we verify that the behavior of individual spin chains in the covered field range is indeed dominated by the longitudinal IC spin fluctuations, one of the ingredients for the concept of tunable interaction. Gapless fluctuations are directly probed by $T_1^{-1}$~\cite{Moriya_1956}. At $2.4$~K, in the TLL state, the $T_1^{-1}$ data precisely follow the decreasing trend of the TLL prediction for the longitudinal IC fluctuations [Fig.~\ref{fig4}(c)]. There are no traces of the transverse staggered fluctuations, which are expected to become important at higher fields~\cite{Okunishi_2007}. The necessary condition for the dominance of the longitudinal IC fluctuations is the anisotropy of the involved $S=1/2$ spins~\cite{Okunishi_2007}. An effective description of this type indeed applies to Co$^{2+}$ spins in trigonally distorted octahedral environments~\cite{Kimura_2006}, the case encountered in BaCo$_2$V$_2$O$_8$. Related Co-based spin-chain systems show promise for use as simulators of frustrated magnetism in other planar geometries, such as triangular or kagome, hence allowing new insights into spin-liquid physics~\cite{Balents_2010}.

In summary, we identified a mechanism for the tunable exchange interaction between the NN spin chains in BaCo$_2$V$_2$O$_8$. This mechanism is responsible for the transition to a new IC ordered state that we discovered, and allows to regard BaCo$_2$V$_2$O$_8$ as a tunable simulator of frustrated magnetism on a square lattice. The novel phenomenon of tunable interaction emerges in the pairs of 1D objects dominated by tunable quantum fluctuations in the presence of frustration. It adds to the list of other emergent phenomena arising from particular types of spin fluctuations in frustrated spin systems, such as emergent gauge fields in classical spin ices or fractional particle excitations in quantum spin liquids~\cite{Balents_2010}.

We thank S.~Kr\"amer, T.~Giamarchi, M.~Grbi\'c, \mbox{M.-H.}~Julien, P.~Jegli\v{c}, R.~\v{Z}itko, A.~Miheli\v{c}, I.~Mu\v{s}evi\v{c} and P.~McGuiness for helpful suggestions. The work was partly supported by the Slovenian ARRS project No. J1-2118, by the French ANR project NEMSICOM, and by the EuroMagNET II network under the EU contract number 228043. M.~K. and C.~B. dedicate the paper to the memory of Prof. R.~Blinc.

\par\clearpage
\section{Supplemental material}

{\bf Experimental details.} We grew a high-quality single crystal of BaCo$_2$V$_2$O$_8$, shaped as a $5$~cm long rod of $3$~mm diameter, using the floating zone technique in a mirror furnace under air. From the rod, a sample with dimensions $2\times 0.8\times 0.2$ mm$^3$ was cut, so that the $c$ axis was perpendicular to the biggest face. Nuclear magnetic resonance (NMR) experiments were performed on $^{51}$V nuclei with spin $I=7/2$ in a magnet that enabled field sweeps up to $15$~T. A copper NMR coil with the sample was installed into the mixing chamber of a dilution refrigerator, so that the $c$ axis of the sample was parallel to the field. $^{51}$V spectra were measured by sweeping the frequency at a fixed magnetic field and summing the Fourier transforms of the echo. To be able to cover wide frequency and field ranges, we used the ``top tuning'' scheme, where a variable tuning delay line and a matching capacitor were mounted outside the NMR probe. The obtained NMR spectra were corrected for the demagnetizing effect by means of the known static magnetization curve. The field was calibrated by $^{63}$Cu NMR of metallic copper in the NMR coil, and the temperature was measured using a calibrated RuO$_2$ resistor. The longitudinal (i.e., $T_1$) relaxation curves were obtained from the integrated spin-echo intensity after the variable time delay following an inversion pulse. Due to the small quadrupole splitting, amounting to $30$~kHz, all nuclear transitions could be irradiated at once, so that the relaxation curves could be reproduced by a simple exponential function. In the ordered phase, we needed to add the stretching exponent. The transverse (i.e., $T_2$) relaxation curves were obtained from the integrated spin-echo intensity with variable time delay between the two pulses. Characteristic quadrupole oscillations prevented us from fitting the curves with the standard decay functions, so that the $T_2$ values were estimated as the delays at which the intensity reached $1/e=0.37$ of its maximum value. The details of magnetic ordering were extracted from the $^{51}$V NMR spectra using the tensors of the hyperfine and dipolar couplings between the $^{51}$V nucleus and the Co$^{2+}$ spins. The hyperfine tensors were determined on the basis of our high-temperature study of orientation-dependent $^{51}$V NMR spectrum, while the dipolar tensors were calculated numerically from the known crystal structure.

\end{document}